\documentclass[aps,twocolumn,floatfix,showpacs,superscriptaddress,prb,eqsecnum]{revtex4}
\usepackage{graphicx}
\usepackage{amsmath}
\usepackage{bm}

\begin{document}
\title{Piezoconductivity of gated suspended graphene}
\author{M. V. Medvedyeva}
\affiliation{Instituut-Lorentz, Universiteit Leiden, P.O. Box 9506, 2300 RA Leiden, The Netherlands}
\affiliation{Kavli Institute of NanoScience, Delft University of Technology, Lerentzweg 1, 2628 CJ Delft, The Netherlands}
\author{Ya. M. Blanter}
\affiliation{Kavli Institute of NanoScience, Delft University of Technology, Lerentzweg 1, 2628 CJ Delft, The Netherlands}

\begin{abstract}
We investigate the conductivity of graphene sheet deformed over a gate. The effect
of the deformation on the conductivity is twofold: The lattice distortion can be
represented as pseudovector potential in the Dirac equation formalism, whereas the
gate causes inhomogeneous density redistribution. We use the elasticity theory to
find the profile of the graphene sheet and then evaluate the conductivity by means
of the transfer matrix approach. We find that the two effects provide functionally
different contributions to the conductivity. For small deformations and not too high
residual stress the correction due to the charge redistribution dominates and leads
to the enhancement of the conductivity. For stronger deformations, the effect of the
lattice distortion becomes more important and eventually leads to the suppression of
the conductivity. We consider homogeneous as well as local deformation. We also
suggest that the effect of the charge redistribution can be best measured in a setup
containing two gates, one fixing the overall charge density and another one
deforming graphene locally.

\end{abstract}

\maketitle
\section{Introduction}
Graphene is a novel material with highly unusual electron properties,
related to the Dirac form of its energy spectrum at low energies, and
demonstrated in many seminal experiements (for review see~\cite{BeenakkerR,CastroNeto}). Experiments on
single-layer graphene have been performed on the flakes obtained
by exfoliation as well as grown on a substrate.

Graphene also has excellent mechanical properties. Indeed, the
elastic properties have been measured on suspended graphene flakes
mechanically deposited over a hole by indentation in an atomic
force microscope~\cite{Changgu,Delftexp,FrankAFM};
the results showed that graphene is incredibly stiff, with the
breaking strength of the order of $40$N/m, the Young modulus of
1~TPa, and possibility to be stretched elastically up to 20$\%$.
Bending properties have been determined experimentally for
several-layer graphene flakes~\cite{Delftexp} and are not
yet available for a monolayer. Theoretically, these properties have been predicted from
the calculations using the
analytical form of the interatomic potential, and from molecular
dynamic studies~\cite{Huang,Huang2}. Graphene is currently one of the most prospective
candidates for high-frequency nanomechanical
resonators~\cite{Bunch,Chen}, with the
quality factor and eigenfrequency extracted from measurements to
be $Q\approx75$, $f_0=70.5$~MHz for a monolayer, and $Q\approx120$,
$f_0=42$~MHz for 15nm thick graphite. Quality factor further increases
with decreasing temperature. An alternative method to investigate
elastic properties of graphene is to put the film on a flexible
substrate and deform the substrate~\cite{grow3,Mohiuddin}.
The strain influences optical phonon spectum~\cite{Mohiuddin},
which has been measured by Raman spectroscopy.

Recent experiments combine mechanical and electrical properties of
graphene by measuring
conductivity~\cite{Bolotin1,Bolotin2,Lau,suspended} of suspended graphene flakes.
This is a very promising
direction since suspended graphene flakes exhibit much higher
mobility than graphene on substrate due to much weaker disorder
~\cite{Bolotin1,Bolotin2}. Potentially electrons can
produce back-action on the resonator. Graphene resonators are
expected to have high sensitivity to mass and prebuilt
strain~\cite{Chen}, so that they can be used to ultra-sensitive
mass detection.

Suspension of graphene flakes always leads to their deformation,
which in turn affects the conduction properties of graphene.
Deformation creates inhomogeneous elongation of the lattice
constant~\cite{CastroNetoD,Katsnelson} which locally affects the
electron spectrum of graphene.
One way to look at the variations of the band structure of the
strained graphene is to perform density functional
calculations~\cite{CastroNetomol}. Alternatively, the variation of
the lattice constant can be represented at the level of Dirac
equation in the form of pseudomagnetic fields~\cite{Ando}. Ref.~
\onlinecite{Katsnelson} pointed out that local shifts of the Fermi
surface in suspended graphene in the vicinity of the Dirac point
can block the conductivity --- if the Fermi-surfaces at different
parts of the flake do not overlap, the conduction is tunnel rather
than metallic. Effects of disorder due to charged impurities and
midgap states, optical and acoustic phonons were taking into
account for calculating conductivity of gated graphene
in~\cite{Stauber}.
For strong enough deformation, graphene quasiparticles can become
localized~\cite{Ahn}.

In experiments, graphene flakes are typically suspended over a
back-gate. This gate redistributes the electron density in the
flake due to the spatial variation of the capacitance.
The regions in the center of
the suspended part of the flake have higher electron density then the regions near the clamping edges, as the central part is closer to the gate. This
density redistribution affects the transmission coefficients
through the entire flake. The corresponding effect on the
piezoresistivity in ballistic regime is of the first order in the
maximum deformation of the flake in the transverse direction, and it increases the conductivity.
This
has to be contrasted with the effect of the pseudomagnetic fields which
suppress the conductivity. The contribution from pseudovector potential depends
on the strain~\cite{Katsnelson} over the flake
and is of the second order
in the maximum deformation. Thus, this contribution is expected to be weaker than effect from the charge
redistribution. We will show however that this effect can be important for graphene under high enough residual stress.
Inhomogeneous deformation of graphene yields the corrections to the conductivity which are of the fourth order in the maximum deformation,
which is even smaller.

In this Article, we calculate the effect of the gate-induced density
redistribution on the conductivity of the graphene
flake.
We find that, indeed, for high residual stress the correction resulting from the pseudovector potential is important, and the correction to the conductivity is negative. We mostly focus on the regime of low residual stress and show that the correction from the charge redistribution becomes the most important.

Experimentally, influence of deformation on the conductivity would be difficult
to observe on a suspended graphene flake
with one gate since the main effect of the gate is the global
shift of the density rather than its redistribution. To separate
density redistribution and elastic deformation, one needs to employ
two gates. For instance, one can use the configuration with a
large bottom gate and a narrow top gate. The bottom gate deforms
the graphene flake and determines the maximum transverse
deformation $\xi_{max}$. When voltage is switched on the narrow
top gate it does not influence much of deformation of the flake
depleting the charge density below the top gate. Since the region
under the top gate has the lowest density it determines the
conductivity of the whole flake. If this region is brought to the
Dirac point, the correction to conductivity is determined only by
the deformation of graphene~\cite{Katsnelson} and is proportional to
$\left(\xi_{max}(V_g)/L\right)^2$, with $V_g$ and $L$
being the voltage applied to the bottom gate and the length of the
strip under the top gate. However, for higher voltages the charge
redistribution is more important, and the correction to conductivity
is proportional to $\xi_{max}(V_g)/d$, $d$ being the distance
to the bottom gate. Instead of the top gate, one can use an AFM tip.

The paper is organized as follows. In Section~\ref{sec:deformation}
we derive equations for the deformation of suspended
graphene from general theory of elasticity. We consider two
situations --- graphene deformed homogeneously by a gate and
graphene deformed locally by an AFM tip. The capacitance between
the gate and suspended graphene varies due to deformation of the
flake. We calculate the density redistribution over the flake
taking into account the shape of the flake. In Section~\ref{sec:piezo}, we use
these results to evaluate correction to the conductivity. We use the perturbation theory to calculate
the transmission eigenvalues, and the correction to the conductivity is obtained using
the Landauer formula. This correction can be big for sufficiently
strong deformations of the flake which can be produced by an AFM
tip. In Section \ref{sec:discussion} we discuss the results and the regimes not considered in this Article.

\section{Deformation of the graphene sheet} \label{sec:deformation}

\begin{figure}[tb]
\centerline{\includegraphics[width=\linewidth]{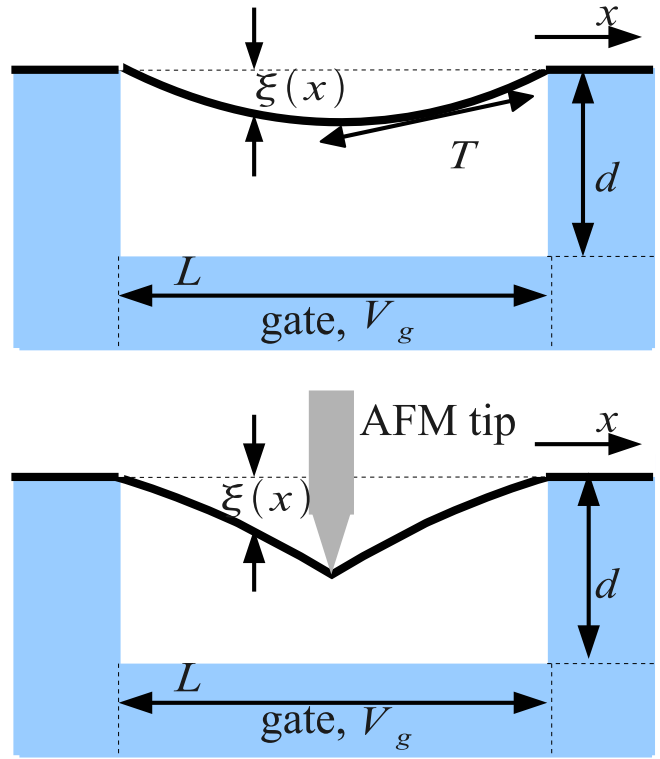}}
\caption{Side view of a deformed graphene flake suspended over a gate.
The deformation is caused only by the interaction with a gate (top) or by the gate and an AFM-tip (bottom).
}
\label{fig1}
\end{figure}

In this Section, we calculate the profile of the graphene sheet formed
by electrostatic forces induced by the gates. For this purpose, we
decompose the total energy of the flake as the sum of electrostatic
and elastic energies. We consider a graphene flake of the length $L$
(direction $x$) and the width $W$ (direction $y$). For simplicity, we
assume $W \gg L$. An undeformed
sheet occupies a part of the plain $z=0$; the electrostatically
induced deflection is $\xi(x,y)$. Below we only consider
small deformations so that we can stay
within the limits of linear theory of elasticity (Hooke's law). At stronger
deformations, as expected from the general theory~\cite{Landau} and
also confirmed by theoretical modeling~\cite{Isacsson} and by
experiments~\cite{Bunch} on graphene, non-linear terms
become important. However, there is a considerable parameter
range, with the displacements up to 50 nm, where the linear regime
is still valid. We discuss the terms which go beyond Hooke's law~\cite{Isacsson}
in Section~\ref{sec:discussion}.

For electrostatic energy, similarly to Ref.~\onlinecite{Blanter}, we
model the system as a capacitor between the flake and the gate, with the
distributed capacitance $C_g$ dependent on the profile of the
flake,
\begin{equation} C_g=\int c[\xi(x,y)] dx dy.\end{equation}
Electrostatic coupling to the leads is modelled via contact
capacitances $C_L$, $C_R$ and resistances $R_L$, $R_R$,
see. Fig~(\ref{fig1}). The total
electrostatic energy of the system carrying the charge $Q$ is
\begin{multline}F_{electr}=-\frac{Q^2}{2C_0}+\frac{Q}{C_0}(C_L V_L +
  C_g V_g)-\frac{C_L C_g V_L V_g}{2C_0}+\nonumber\\
  +\frac{C_L V_L^2 (C_0-C_L)}{2C_0}+
+\frac{C_g V_g^2 (C_0-C_g)}{2 C_0} \
, \label{energy_electr}\end{multline}

with $C_0=C_L + C_R + C_g$.

From now on, we assume that the contacts are ideal,
$C_L=C_R=0$, and thus the electrostatic energy is \begin{equation} \label{elstatenergy}
F_{electr}=-\frac{Q^2}{2C_g}+Q V_g.\end{equation}
The effect of non-ideal contacts is discussed in Section
\ref{sec:discussion}.

\subsection{Elastic energy}

We evaluate the elastic energy in the thin-plate approximation.
The elastic energy consists of the bending contribution
$F_1(\xi(x,y))$ and the stretching contribution $F_2(u_{\alpha
  \beta}(x,y))$, where $u_{\alpha\beta}(x,y)$ is the deformation
tensor, and $\alpha$ and $\beta$ denote the
coordinates in the plane of the sheet ($x$ and $y$). In the linear
regime, the bending contribution is less important than the
stretching one, however, we consider both contributions for
completeness. Explicitly, we have~\cite{Landau}
\begin{multline} F_1(\xi)=\frac{D}{2}\int\int dx dy
\left(\frac{\partial^2\xi}{\partial x^2}+\frac{\partial^2\xi}{\partial y^2}\right)^2+\\
+\int\int dx dy (1-\sigma^2)\left[\left(\frac{\partial^2\xi}{\partial x \partial
    y}\right)^2-\frac{\partial^2\xi}{\partial x^2}\frac{\partial^2\xi}{\partial
    y^2}\right] \label{bending}
\end{multline}
and
\begin{equation} F_2(u_{\alpha \beta})=h_0\frac{u_{\alpha
      \beta}\sigma_{\alpha \beta}}{2}\label{stretching} \ . \end{equation}
Here $D=Eh_0^3/(12 (1-\sigma^2))$ is the bending ridgity, $E$ is the Young
modulus, $\sigma$ is the Poisson ratio, $h_0$ is the thickness of the plate (graphene flake) , and $\sigma_{\alpha\beta}$  is the stress tensor.

In addition, if a local force (for instance, an AFM tip) acts on the
graphene flake, it is best represented by external pressure
$P_{ext}(x,y)$. The work of this external pressure to deform the flake
by $\delta\xi(x,y)$ is  $F_3 = \int P_{ext} \delta \xi(x,y) df$, where
$df$ is the surface element.

The profile of the sheet is determined by minimizing its total
energy. Performing the variation, we find the equation describing the
shape of the flake,
\begin{eqnarray}
  D \Delta^2 \xi-\frac{\partial}{\partial
  x_\beta}(h_0\sigma_{\alpha\beta}\frac{\partial \xi}{\partial
  x_\alpha}) =  P_{el}(x,y)+P_{ext}(x,y),\label{generalform}\\
  \frac{\partial \sigma_{\alpha \beta}}{\partial x_\beta} =  0,\label{generalstretch} \end{eqnarray}
%\begin{equation}
%  D \Delta^2 \xi-\frac{\partial}{\partial
%  x_\beta}(h_0\sigma_{\alpha\beta}\frac{\partial \xi}{\partial
%  x_\alpha}) =  P_{el}(x,y)+P_{ext}(x,y),\label{generalform}\end{equation}
%  \begin{equation} \frac{\partial \sigma_{\alpha \beta}}{\partial x_\beta} =  0,\label{generalstretch} \end{equation}
%\begin{eqnarray} \label{elasteq} D \Delta^2 \xi-\frac{\partial
%}{\partial
%  x_\beta}(h_0\sigma_{\alpha\beta}\frac{\partial \xi}{\partial
%  x_\alpha}) & = & P_{el}(x,y)+P_{ext}(x,y),\label{generalform}\\
%\frac{\partial \sigma_{\alpha \beta}}{\partial x_\beta} & = & 0,\label{generalstretch} \end{eqnarray}
with $P_{el}(x,y)$ being the electrostatic pressure on the plate,
induced by the variation of electrostatic energy (\ref{elstatenergy}).
For ideal contacts, $P_{el}(x,y)=n^2(x,y)/2\epsilon_0$. Here
$n(x,y)$ is the electron density. Eq. (\ref{generalform}) is the most general
equation for $\xi(x,y)$ in the linear approximation of theory
elasticity. For an infinitely wide graphene flake, $W \gg L$, the
deformation in the $y$ direction is homogeneous.

At sufficiently small deformations, the tension along the sheet is
constant over the sheet~(\ref{generalstretch}),
$h_0\sigma_{\alpha\beta}^0=T\delta_{\alpha\beta}$. The tension $T$ is the
sum of two contributions:
\begin{equation} T=T_0+T_H,~T_H=\frac{Eh_0}{1-\sigma^2}\Delta L/L\label{tension}\end{equation}
The first one, $T_0$, is the residual stress
which results from the fabrication process or is induced by the
ripple formation~\cite{Delftexp,Changgu}. The second
contribution, $T_H$, is an internal force due to the relative elongation $\Delta L/L$ (Hooke's law). If we take this term into account,
we can go beyond the thin-plate approximation and consider deformations bigger
then the thickness of the graphene layer.

In the two following Subsections, we solve the above equations for
two specific situations: homogeneous external force (which can be produced by a bulk bottom gate),
and local force (produced for example by an AFM tip).

\subsection{Homogeneous force: Deformation by a bottom gate}

Applying a voltage on a bottom gate is a standard way to vary
electron density in graphene. If the suspended graphene flake is
charged, it is subject to a mechanical force proportional to the charge density. If the area of
the gate is much larger than the area of the flake, the electron
density induced by the gate is constant almost everywhere, $n=Q/WL$,
except for the clamping points of the flake, where it is determined
not only by the solution of the Poisson equation (providing
singularities at the capacitor edges), but also by the metallic leads
to which the flake is clamped. 
Indeed, experimental evidence for this
charge inhomogeneity exist and can be accessed by asymmetry of the Dirac
peak in conductivity~\cite{smth}. However, these density inhomogeneities
at the clamping areas very little affect the deformation, since the
displacement vanishes at the edges of the flake. Therefore we can
approximate the effect of the gate by homogeneous
electrostatic pressure over the flake, $P=\epsilon_0 V_g^2/ 2 d^2$,
$V_g$ and $d$ being the gate voltage and the distance to the gate. The profile of the graphene sheet
is found from the equation
\begin{equation} D\frac{\partial^4 \xi}{\partial x^4}-T\frac{\partial^2 \xi}{\partial x^2}=P,\end{equation}
where the stress $T$ is constant over the sheet~(\ref{tension}) and
the deformation-dependent contribution to it depending has to be found
self-consistently,
\begin{equation} \label{induced_stress}
T_H = \frac{Eh_0}{2(1-\sigma^2)}\int_0^L \xi'^2(x) dx
\end{equation}
(the case for inhomogeneous $T_H$ derived
in Ref.~\onlinecite{Isacsson} is discussed in Section~\ref{sec:discussion}
and does not induce significant difference in results).
For the boundary condition corresponding to the
clamping the sheet,
$\xi(0)=\xi(L)=\xi'(0)=\xi'(L)=0$, the profile is
\begin{eqnarray} \xi(x)& =& \frac{PL}{2T\mu}\left[\frac{\sinh \mu L}{\cosh
\mu L -1}(\cosh \mu x -1 ) - \sinh \mu x +\mu x \right. \nonumber \\
& - & \left. \frac{\mu x^2}{L} \right], \ \ \ \mu
  =\sqrt{\frac{T}{D}}\ .
\label{flakeform}\end{eqnarray}
The profile~(\ref{flakeform}) is parabolic in the middle of the strip
(as noted in Ref. \onlinecite{Katsnelson}). As we show below, in
graphene the dimensionless parameter $\mu L$ assumes large values. In
this case, the profile can be simplified, and near the
middle of the strip has the form
\begin{equation} \xi(x)=\frac{PL}{2T}\left(x -\frac{x^2}{L} \right) \label{formmy} .\end{equation}
Close to the edges, the profile becomes $\xi(x)=P\mu L x^2/4T$. Substituting this shape into Eqs. (\ref{bending}) and (\ref{stretching}), we find the values of the parameters $F_1$ and $F_2$,
\begin{equation} F_1=\frac{P^2 LW}{16 T \mu^2(-8+\mu L)},\end{equation}
and
\begin{equation} F_2=\frac{T^2 WL (1-\sigma^2)}{Eh_0}.\end{equation}
The maximum
vertical displacement obeys the equation
\begin{equation} \xi_{max}=\frac{PL^2}{8(T_0+8 Eh_0\xi_{max}^2/(3(1-\sigma^2)L^2))} \label{ximax}. \end{equation}

\begin{figure}[t!]
\centerline{\includegraphics[width=\linewidth]{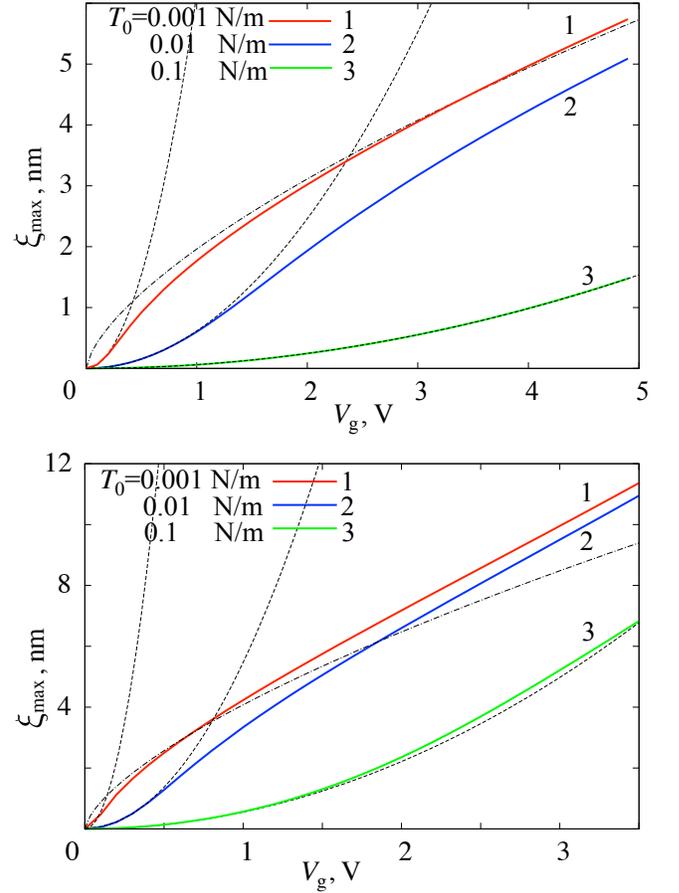}}
\caption{Dependence of the maximum deviation on the gate
  voltage, $\xi_{max}(V_g)$. The solid curves represent the self-consistent solution of nonlinear coupled equations for the deformation of the flake and the charge induced by the gate, Eqs.~(\ref{ximax}) and (\ref{charge}). The distance to the gate is $d=300$~nm (top panel) and $d=100$~nm (bottom panel).
  Other parameters of the graphene flake, length $L=1~\mu$m, Young's modulus $E=1$~TPa, Poisson ratio $\sigma=0.15$, and the thickness of the flake $h_0=0.34$~nm, are chosen in order to model real experimental data.
  The results are given for the different values of the residual stress: the curve $1$ is for $T_0=0.001$~N/m, $2$ is for $T_0=0.01$~N/m, $3$ is for $T_0=0.1$~N/m. For each value of the residual stress, the asymptotic curves at low gate voltages~(\ref{larget0}) are shown as dashed lines, the curve $3$ for high residual stress coincide perfectly with its assymptote. The asymptotic curves for low residual stress, Eq.~(\ref{smallt0}), are shown by dashed-dotted lines. The correspondence between the solution of equations and asymptotics for low residual stress is not perfect. The reason is that the asymptotics are calculated for the linear charge-voltage dependence, and $n(V_g)$ is non-linear according to Eq.~(\ref{charge}) for sufficiently high gate voltages on the flake.}
\label{chi}
\end{figure}

The deformation of the sheet leads to the redistribution of the
electron density, which in the Thomas-Fermi approximation is $n(x) =
V_g \epsilon_0/(d-\xi(x))$. In its turn, the density redistribution affects
the profile of the sheet, and needs, in principle, to be
calculated self-consistently. However, as soon as the displacement $\xi_{max}$ is much
smaller than the distance to the gate, the later effect is
insignificant (of the order of $\xi_{max}/d$),
and we will use the shape (\ref{flakeform}) not modified by the
density redistribution.

The charge over the graphene flake is determined from minimization of the total energy of
the system with respect to electron density $n$,
\begin{equation} -V_g+\frac{nd}{\epsilon_0}\left[ 1-\frac{8}{3}\frac{\xi_{max}}{d}+\left(1+\frac{1}{2}\right)\frac{\xi_{max}}{d\mu L}+\frac{1}{3}\frac{\xi_{max}}{d}\right]=0\label{charge0},\end{equation}
where the maximum deformation of the sheet in the middle, $\xi_{max}$ (\ref{ximax}),
and depends on charge density.
The second term in the brackets and $1$ from the third term come from electrostatic energy and originate from the
redistribution of the charge density due to variation of the distance between parts of deformed graphene and the gate.
The rest ($1/2$) of the third term comes from bending energy.
The fourth term takes into account dependence of the stretching force $T$ over the flake
on the charge density via the deflection (Eq. (\ref{induced_stress})).
Calculations are made under the assumption $\mu L \gg 1$,
which is realistic for available experiments. Simplifying Eq. (\ref{charge0}), we obtain
\begin{equation} -V_g+\frac{nd}{\epsilon_0}\left(1-\frac{7}{3}\frac{\xi_{max}}{d}+\frac{3}{2}\frac{\xi_{max}}{d\mu L}\right)=0\label{charge}\end{equation}
At low gate voltages, Eq. (\ref{charge}) yields the linear gate
voltage dependence of the electron density, $n_0 \equiv V_g\epsilon_0/d$.
There is non-linear deviation from this dependence at higher gate voltages and at rather low
initial strain $T_0$.

The maximum deformation can be expressed analytically in two limiting
cases. First, if the residual stress $T_0$ is stronger than the
induced stress $T_H$, it mostly accounts for the deformation of the sheet,
\begin{eqnarray} T_H & = & \frac{Eh_0P^2L^2}{24(1-\sigma^2) T_0^2} \ll T_0,
\left( \frac{Eh_0 P^2 L^2}{24(1-\sigma^2)} \right)^{1/3}\ll T_0, \nonumber \\
  \xi_{max} & = & \frac{\epsilon_0 V_g^2 L^2}{16 d ^2 T_0}
  \ , \label{larget0}\\
  \frac{n-n_0}{n_0}& = & \frac{7}{3} \frac{\xi_{max}}{d}\left(1-\frac{72}{7\sqrt{T_0/D}L}\right).\label{larget0charge}\end{eqnarray}

In the case of low residual stress, one obtains
\begin{eqnarray} T_0\ll T_H & = &\frac{1}{2} \left( \frac{Eh_0P^2L^2}{3(1-\sigma^2)} \right)^{1/3},
\nonumber \\
  \xi_{max} & = &
  \frac{1}{4}\left(\frac{3V_g^2\epsilon_0(1-\sigma^2)L^4}{2d^2 Eh_0}\right)^{1/3}, \label{smallt0}\\
  \frac{n-n_0}{n_0} & = & \frac{7}{3} \frac{\xi_{max}}{d}-12\sqrt{\frac{3D (1-\sigma^2)}{2Eh_0}}. \label{smallt0charge}
\end{eqnarray}

The maximum deviation~$\xi_{max}$, obtained from the numerical solution of
coupled nonlinear equations Eqs.~(\ref{ximax}), (\ref{charge}), as well as asymptotic expressions (\ref{larget0}) and (\ref{smallt0}),
are shown in Fig.~\ref{chi} for different values of initial stress $T_0$~\cite{foot1}.
According to Eqs.~(\ref{larget0charge})~and~(\ref{smallt0charge}), the nonlinear part of the charge induced on
the graphene flake follows the dependence $\xi_{max}(V)/d$.
\begin{figure}[tb]
\centerline{\includegraphics[width=\linewidth]{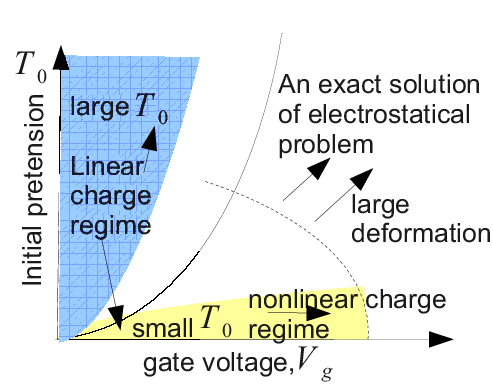}}
\caption{Regimes of the deformation of suspended graphene.
For large residual stress $T_0$ the asymptotics ~Eq.~(\ref{larget0}) are valid, and the charge on the flake follows the gate voltage linearly.
For small $T_0$ the asymptotics ~Eq.~(\ref{smallt0}) are valid, the charge is linear with the gate voltage at low gate voltages and starts to follow non-linear dependence with increasing voltage.
At high gate voltages, when the deformation of the flake is bigger than ~$\xi_{max}/d\sim0.1$, one needs to solve self-consistently the electrostatic problem and the problem of elasticity, analytical results for this region are not available.
}
\label{regimes}
\end{figure}
Consequently, we encounter several regimes for the deformation,
\begin{itemize}
\item at large $T_0$ the charge--voltage dependence is linear for realistic parameters because the maximum deformation is not too large for realistic characteristics of graphene flake. It is shown in Fig.~\ref{chi} for $T_0=0.1$~N/m that the maximum deformation is in a good agreement with Eq.~(\ref{larget0});
\item at small $T_0$ and low gate voltages $V_g$ the charge--voltage dependence can be in linear regime, and the maximum deformation follows Eq.~(\ref{smallt0}). We illustrate this for the flake with the parameters $T_0=0.001$~N/m and distance to the gate $d=300$~nm (See. Fig.~\ref{chi}, top), where the solutions of coupled electrostatic and elastic equations, Eq.~(\ref{ximax}), (\ref{charge}), follow asymptotic expression Eq.~(\ref{smallt0}). The charge redistribution does not need to be taken into account;
\item at small $T_0$ and large $V_g$ the system is in the non-linear charge regime. This situation can be realized for small distances to the gate when the coupling of the graphene sheet to the gate is large, so that it is possible to create large deformations using low gate voltages. For example, at $d=100$~nm the non-linear charge regime influences the deformation already at voltage $V_g=2$~V, at the bottom plot Fig.~\ref{chi} we can see the intersection of the asymptotical curve (\ref{smallt0}) and the actual solution of Eqs.~(\ref{ximax}), (\ref{charge}).
\end{itemize}
The schematic representation of these regimes is shown in Fig.~\ref{regimes}.

\subsection{Local force: Deformation by an AFM tip}

Next, we consider a concentrated force acting on graphene. This force can be provided,
for example, by an AFM tip. The effect of the tip
is modeled by strong pressure exerted on a narrow area of the width $l
\ll L$. We assume that the problem is still homogeneous in the
$y$-direction, what simplifies the calculations enormously. Inclusion
of a pressure action in a narrow circle, which is experimentally
relevant for an AFM tip, is not expected to bring qualitatively new
features. We consider pressure $P(x)=P_2, 0<x<L/2-l/2, L/2+l/2<x<L$ and $P(x)=P_1,L/2-l/2<x<L/2+l/2$. Here
$P_2$ is the local pressure, and $P_1 \ll P_2$ can describe homogenious pressure due to electrostatics.

The maximum displacement of the flake (realized at the central point)
is easy to write down for $\mu L \gg 1$ and $l \ll L$:
\begin{eqnarray} \xi_{max}&=&\frac{P_1\left((\mu L/2)^2 e^{\mu l/2}-2 e^{-\mu l}\right)}{2\mu^2T}+\nonumber\\
&+&\frac{ P_2\left(e^{-\mu l}+e^{\mu l}\mu^2 l L/4  \right)
}{\mu^2 T}.\label{xiafm}\end{eqnarray}
For $P_1\ll P_2$ and $1\ll e^{\mu l/2} L\mu l\mu/4$ we obtain
\begin{equation} \xi_{max}=\frac{P_2 l L}{4T}.\end{equation}

The profile of the graphene sheet in this approximation becomes
\begin{equation} \xi(x)=\frac{2 \xi_{max}}{L}|x-L/2|. \label{localprofile}\end{equation}

In the limits of weak and strong residual stress the deformation is determined by
\begin{equation} T_H = \frac{Eh_0}{8(1-\sigma^2)}\left(\frac{P_2 l}{T_0}\right)^2 \ll T_0,
\left(\frac{Eh_0}{8(1-\sigma^2)}P_2^2 l^2\right)^{1/3}\ll T_0, \nonumber \end{equation}
\begin{equation}  \xi_{max}=\frac{P_2 l L}{4 T_0}; \label{largep2}
\end{equation}
and
\begin{equation} T_0\ll T_H  = \frac{1}{2}\left(\frac{Eh_0 P_2^2 l^2}{1-\sigma^2}\right)^{1/3},
  T_0\ll \frac{1}{2}\left(\frac{Eh_0 P_2^2 l^2}{1-\sigma^2}\right)^{1/3},\nonumber \end{equation}
 \begin{equation} \xi_{max}=L\left(\frac{P_2 l (1-\sigma^2)}{8Eh_0}\right)^{1/3}. \label{smallp2}
\end{equation}

\begin{figure}[tb]
\centerline{\includegraphics[width=\linewidth]{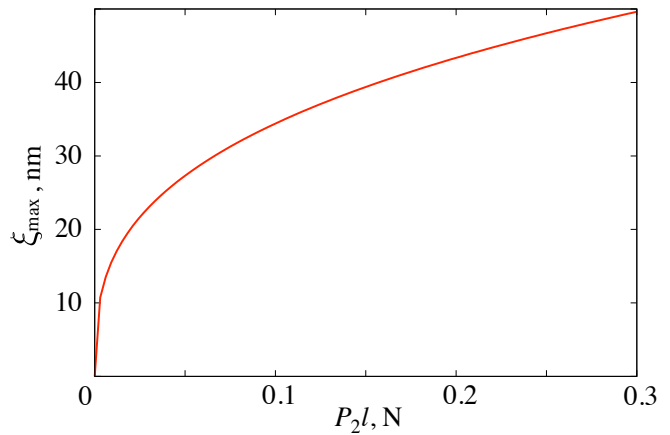}}
\caption{Dependence of the maximum deformation $\xi_{max}$ on the applied force $P_2 l$ for the case of the point force in the middle of the graphene sheet. Only the curve for $T_0 = 0$ is shown, since the residual stress is not important for this case: The strain created by deformation becomes large (more than $0.1$~N/m) already at moderate deformations in the middle, $\xi_{max}\sim10$~nm. Other parameters of the flake are~\cite{foot1}.}
\label{ximaxpoint}
\end{figure}

The dependence of the maximum deformation on the applied external local force is shown in Fig.~\ref{ximaxpoint}.
The deformation produced by this force is much bigger than the deformation caused by electrostatic pressure of
the gate. The electrostatic problem for this case can be solved separately from the problem of elasticity.

\section{Piezoconductivity of graphene flake}
\label{sec:piezo}

It was shown experimentally~\cite{Bolotin2} that suspended graphene flakes are described with good precision as
purely ballistic. Theoretically, conductance is determined by Landauer
formula~\cite{Carlo}
\begin{equation}
\sigma=4e^2/h\sum_{n=0}^{N-1} T_n, \label{Landauer}
\end{equation}
where $T_n$ is the transmission eigenvalue in the transport channel
$n$, and the factor $4e^2/h$ is conductivity of a single transport channel
which takes into account valley and spin degeracy. The number of open
transport channels $N=W k_F/\pi$ is proportional to the Fermi momentum
$k_F = (\pi n/e)^{1/2}$,
and thus the conductivity is proportional to the
square root of the electron density $n$, $\sigma \propto \sqrt{n}$.

The conductivity of graphene flake suspended over a gate can deviate from this
dependence. To start with, due to electrostatic
interaction with the gate, the density becomes inhomogeneous~\cite{Silvestrov}.
In particular,
Poisson equation leads to the square root divergence of the electron
density at the clamping points, as in any capacitor (see
e.g. Ref. \onlinecite{Morse}). To treat this divergence properly, one
has to take into account electrostatic interaction with the
contacts near the edge of the graphene strip, which modifies
significantly the electron density near the edge, removing the
divergence. However, the effect of this inhomogeneous density close to
the contacts does not affect the piezoconductivity of the flake, since
the deformation close to the clamping points is very weak, and thus it
can be included into the contact resistance at the clamping points.

We now turn to the effects of the deformation on the
conductivity. Deformation of graphene can change the conductivity by inducing changes
in the band structure (which results in
pseudo-magnetic fields) as well as by changing the electron density over the
flake. We consider both these mechanisms and will show that typically
the effect of the density redistribution dominates.

Electrons in graphene obey Dirac equation. Deformation of
the flake influences on the Dirac equation in three ways --- it shifts
the K-points by a certain amount $\delta k/k_F$ (pseudomagnetic field), renormalizes
the Fermi velocity by $\delta v_F/v_F$, and induces the variation of
the electron density on the flake $\delta n/n$. The deformation
correction to the conductivity is thus a function of these three
dimensionless parameters.

The pseudomagnetic field, produced by the shift of the $K$-point, is caused by stretching and bending. The shift of the $K$-point due to
stretching generates the vector potential~\cite{Ando,Katsnelson}
\begin{equation}
A_{y}^{str}=\frac{C\tilde{\beta}}{a}(u_{xx}-u_{yy}),\ \ \  A_{x}^{str}=-2\frac{C\tilde{\beta}}{a}u_{xy}, \label{pseudovector}
\end{equation}
where $C$ is the order of 1, and $\tilde{\beta}=-\partial \log (t)/\partial \log(a)$,
$t$ and $a$ being the overlap integral in the tight-binding model and
the lattice parameter, respectively.
For $L \ll W$ one has $u_{xy}=0$,
and hence $A_x^{str}=0$. The deformation is homogeneous within the limits of
applicability of Hooke's law, and thus $u_{xx}={\rm const}$ and
$A_y^{str}={\rm const}$. This means that there is no pseudomagnetic field over the
graphene flake. The pseudomagnetic field only appears in the region where the flake goes from the substrate to the suspended state~\cite{Katsnelson} and, as noted above, its effect to the piezoconductivity is small, of the second order in $\xi_{max}/L$,
\begin{equation} \frac{d\sigma_K}{\sigma}=\frac{A_{y}^{str}}{k_F}, \label{condK1}\end{equation}
where the deformation on the edges has been estimated as $u_{xx}=\xi_{max}^2/L^2+T_0(1-\sigma^2)/Eh_0$, and $k_F=\sqrt{\pi\epsilon_0 V_g/de}$.  Taking into account the value of $C\tilde{\beta}/a$~\cite{Cbeta}, we obtain
\begin{equation}  \frac{\delta \sigma_K}{\sigma}=205 \sqrt{\frac{d[\mu\text{m}]}{V_g[\text{V}]}}\left( \frac{\xi_{max}^2}{L^2}+\frac{T_0(1-\sigma^2)}{Eh_0}\right).\label{condK2}\end{equation}
Note the contribution from two terms induced by deformation stress and residual stress, as well as multiplication with the big prefactor $205$.

The underlying physical picture for the model of Ref.~\onlinecite{Katsnelson} is that the graphene flake is "glued" to the walls at the
suspension point. Whereas this has been realized in some experiments~\cite{Changgu}, it describes the situation when the residual strain $T_0$ is of the same order or higher than the strain induced by the gate voltage. The residual strain results from the fabrication process and is most likely to
be created by impurities in the substrate. It can be made low on purpose since the strain is reduced after annealing~\cite{Chen}. In the opposite situation, when the residual stress is not significant, the pseudomagnetic field is inhomogeneous and distributed over the whole suspension area.

The pseudomagnetic field is also inhomogeneous if one considers the bending contribution. Bending leads to the inhomogeneous modification of
the overlap of the orbitals, and the resulting pseudovector potential has the form~\cite{Katsnelson2}
\begin{equation}
A_{y}^{bend}=\frac{t_{bend}}{a}\left(\frac{\theta^2(a,x)}{2}-
\frac{\theta^2(\frac{a}{2},x)}{2}\right)\ , \end{equation}
$\theta(a,x)$ being the angle between normal vectors to the graphene surface at  the
points $x$ and $x+a$, and the constant~\cite{Kim} $t_{bend}
=3.21$~\cite{tbend}. The shape dependence of $\theta(a,x)$ has the form
\begin{displaymath}
\theta^2(a,x)=a^2\left(\frac{\partial^2 \xi}{\partial
  x^2}\right)^2\left(1+\left(\frac{\partial \xi}{\partial
  x}\right)^2\right)^{-1} \ .
\end{displaymath}
This yields $ A_{y}^{bend}\approx (3t_{bend}/8a)(\xi_{max} a/L^2)^2$. Hence the
contribution from bending is approximately $(a/L)^2$ times smaller then
from stretching without residual stress, and is thus negligibly small, even though the resulting magnetic field is not
homogeneous.

\begin{figure}[tb]
 \centerline{\includegraphics[width=\linewidth]{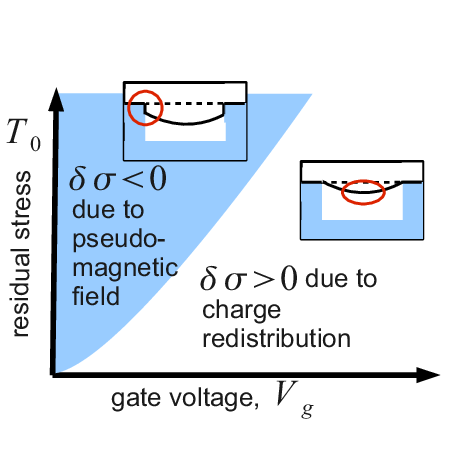}}
 \caption{Schematic behavior of piezoconductivity. For low residual stress $T_0$ the correction is mainly due to the charge redistribution and has positive sign. For high residual stress the correction is negative.}
 \label{regimescond}
\end{figure}

The easiest way to estimate inhomogeneous stretching of graphene is to take Hooke's
law in the local form, $T_H (x)=Eh_0 u_{xx}(x)$. Since the maximum relative deformation can be estimated as
$u_{xx}=\xi_{max}^2/L^2$, naively, the correction from non-homogeneous stretching is of the same order as
the one from delta-functional pseudomagnetic field at the clamping edges. We show below, however, that the correction
from non-uniform stretching is of the order of $\xi_{max}^4/L^4$, but still due to large prefactor it can reduce the
conductivity at low gate voltages.

Another effect induced by the deformation is the renormalization of the Fermi velocity. The renormalized value of the velocity can be derived from the tight-binding model. Assuming that the graphene sheet is only deformed in the $x$-direction, we find that the $x$-component of the Fermi velocity is unchanged wheres the $y$-component is renormalized,
\begin{equation} v_{Fy}=v_F(1-C\tilde{\beta}u_{xx}),\end{equation}
so that approximately $v_{Fy}\approx v_F(1-\xi_{max}^2/L^2)$. The effect of the renormalization on the conductivity is not significant and has the order of magnitude $\xi_{max}^2/L^2$. Note that this is the same dependence on $\xi_{max}/L$ as for pseudomagnetic fields, however, it is not enhanced by a big prefactor.

The influence on conductivity of such change in fermi-velocity is not significant. This influence can be in principe measured experimentally as the conductivity variation at the Dirac point, similarly to how we explain below in Subsection ~\ref{subsec:twogates}.

\subsection{Correction to conductivity due to the charge redistribution}

\begin{figure}[tb]
\centerline{\includegraphics[width=\linewidth]{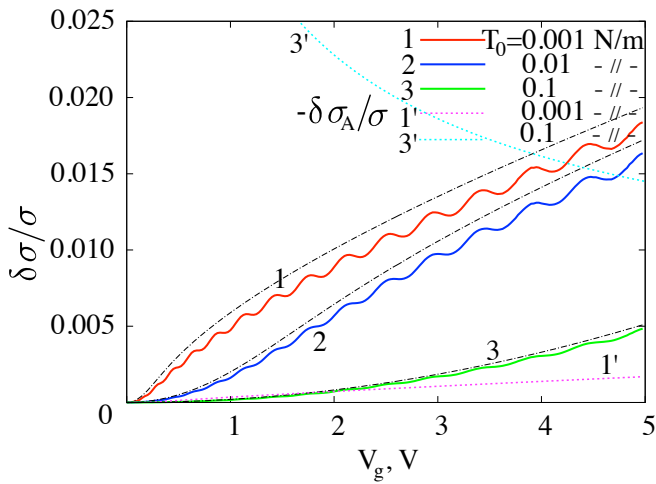}}
\caption{The correction to conductivity. Parameters of the flake are the same as for Fig.~\ref{chi} (top). Asymptotic expressions for the high gate voltage are shown by the dashed-dotted lines. Additionally, the correction due to the delta-functional pseudomagnetic field at the suspension regions~\cite{Katsnelson}, Eq.~(\ref{condK2}), is shown (with the opposite sign).}
\label{correctiongate}
\end{figure}

Redistribution of electric charge due to interactions with the
gate is found from the assumption that the potential along the
graphene sheet is constant, $U(x)=\delta Q(x)/\delta C(x)=const$, where
$\delta C(x)$ is the capacitance of the element of the length $\delta x$ of graphene, $\delta C(x)=W \delta x/4\pi (d-\xi(x))$, and
$\delta Q(x)=n(x) W \delta x$ is the charge of this element.
In the first order approximation,
this gives $\delta n (x)/n_0=\xi(x)/d$.

The conductivity of graphene is proportional to charge density $n$, and thus the contribution to conductivity
due to charge redistribution is expected to be linear in the maximum deviation from the homogeneous density, $\delta n_{max}$.
Thus, the correction to conductivity is expected to be $\delta \sigma/\sigma\sim \xi_{max}/d$.
\begin{figure}[tb]
\centerline{\includegraphics[width=\linewidth]{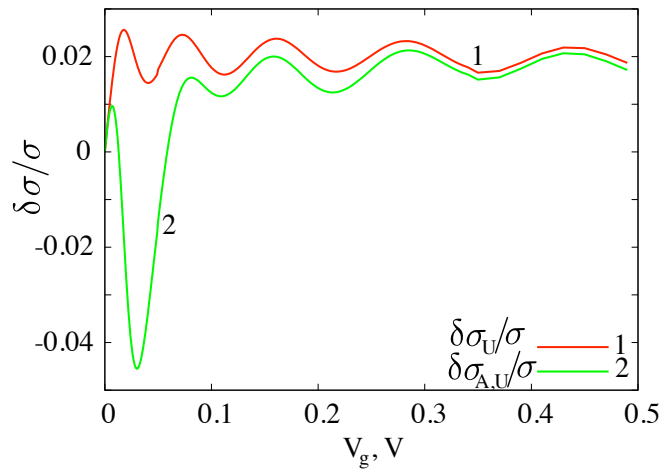}}
\caption{Dependence of the piezocorrection to conductivity on the gate voltage for fixed deformation, obtained by solving the Dirac equation by exact transfer matrix method. Here, $\delta \sigma_U$, the curve marked $1$, is the correction with only charge redistribution taken into account, and $\delta \sigma_{A,U}$, the curve $2$, encompasses both contributions, the one due to non-uniform tension and the one due to charge redistribution. At low gate voltages the correction $\delta \sigma_{A,U}$ is mostly caused by pseudo-magnetic field and is negative, for higher gate voltages it changes sign and approaches to $\delta \sigma_U$. Parameters of the graphene flake are~\cite{foot1}.}
\label{correctionA2}
\end{figure}
Before starting the calculation of the correction to conductivity, we estimate the range where this correction
of the order of $\xi_{max}/d$ is more important than the correction due to pseudovector potential which we considered above, Fig.~\ref{regimescond},
\begin{equation} T_0[\text{N/m}]<10^{-3}\frac{L}{d}\sqrt{\frac{V_g^{5/2}[\text{V}]}{d^{3/2}[\mu\text{m}]}}\label{deformAA}.\end{equation}
As noticed in Ref.~\onlinecite{Katsnelson} the pseudovector potential at low gate voltages blocks conductivity, this is seen from Eq.~(\ref{condK2}).
For large deformation the expressions~(\ref{smallt0}) are valid, residual stress is not important any more, and thus the gate voltage should be large enough to see the decrease of conductivity,
\begin{equation} V_g [\text{V}] >2.8 \frac{L^4}{d^4}\frac{1}{d[\mu\text{m}]}.\end{equation}
This deformation is so strong that in can not be reached in practice.

For the deformation with AFM the residual stress is not important and at deviations
\begin{equation} \xi_{max}[\text{nm}]>2.5 \frac{L^2[\mu\text{m}]}{d[\mu\text{m}]}\sqrt{\frac{V_g [\text{V}]}{d [\mu \text{m}]}} \label{deformA}\end{equation}
correction to pseudovector potential starts to suppress the conductivity.

To calculate the correction due to the charge redistribution,
we notice that the density variation is translated into the correction for
conductivity via the variation of the transmission probabilities $T_n$,
which are the eigenvalues of the matrix $\hat t^{\dagger}\hat t$,
$\hat t$ being the transmission matrix of the graphene sheet. The transmission eigenvalues $t_q$
are determined in Appendix by the transfer matrix method.
The correction to the conductivity
is linear in the density shift $\delta n$, and consequently in the maximum deformation $\xi_{max}$
(as is shown above from simple qualitave considerations). It has
the form (see Appendix)
\begin{eqnarray} \delta\sigma_{U}&=&\sum_q 4 \left\vert t_q \right\vert^2 \frac{q^2 k_F^2}{k^3} \sin kL  \nonumber\\
&\times& \int_{0}^{L} d x \frac{\xi(x)}{d} \sin k(L-x) \sin kx, \label{correctionUmain}
\end{eqnarray}
where $\vert t_q \vert^2 =(\cos^2 kL + k_F^2\sin^2 k L/k^2 )^{-1}$ is the transmission probability for the mode labeled by the
transverse momentum $q=2\pi n/W$, $n$ being an integer number, and $k$ is a wave number in the direction along the strip,
so that $k_F^2=k^2+q^2$.

To carry out more detailed analysis, we consider specific
deformation setups discussed in Section~\ref{sec:deformation} ---
homogeneous and local deformation.

Eq.~(\ref{correctionUmain}) can be analyzed analytically for small and large values of the parameter $k_FL$,
which characterizes the charge density over the flake. The correction to conductivity for the homogeneous deformation (bottom
gate) has the following asymptotic behavior for small and large values of $k_FL$ (for more details, see Appendix),
\begin{equation}
\frac{\delta\sigma}{\sigma} = \begin{cases}
\xi_{max}/2d, & \text{ $1\ll k_FL;$}
\\
0.021\xi_{max}(k_FL)^2/d, & \text {$k_FL\ll
1.$} \end{cases}
\label{deltasigma}
\end{equation}

Taking into account the functional dependence of the maximum deviation for small and large initial stress $T_0$,
Eqs.~(\ref{smallt0}) and (\ref{larget0}), we get the asymptotic dependence of the correction to conductivity on the gate voltage,
for $T_H\ll T_0$:
\begin{equation}
 \frac{\delta\sigma (V)}{\sigma(V)}
 \sim
       \begin{cases}
	  L^2 V^2/T_0 d^3,&\text{ $1\ll k_FL$;}\\
	  L^4 V^3/T_0d^4, &\text{$k_FL \ll 1,$}
\end{cases}
\label{condth}
\end{equation}

and for $T_0\ll T_H$:
\begin{equation}
 \frac{\delta\sigma (V)}{\sigma(V)}
 \sim
 \begin{cases}
  L^{4/3}V^{2/3}/d^{5/3},&\text{$1\ll k_FL;$}\\
  \left(VL^2/d^2\right)^{5/3}, &\text{$k_FL\ll 1.$}
\end{cases} \label{condt0}
\end{equation}
\begin{figure}[tb]
\centerline{\includegraphics[width=\linewidth]{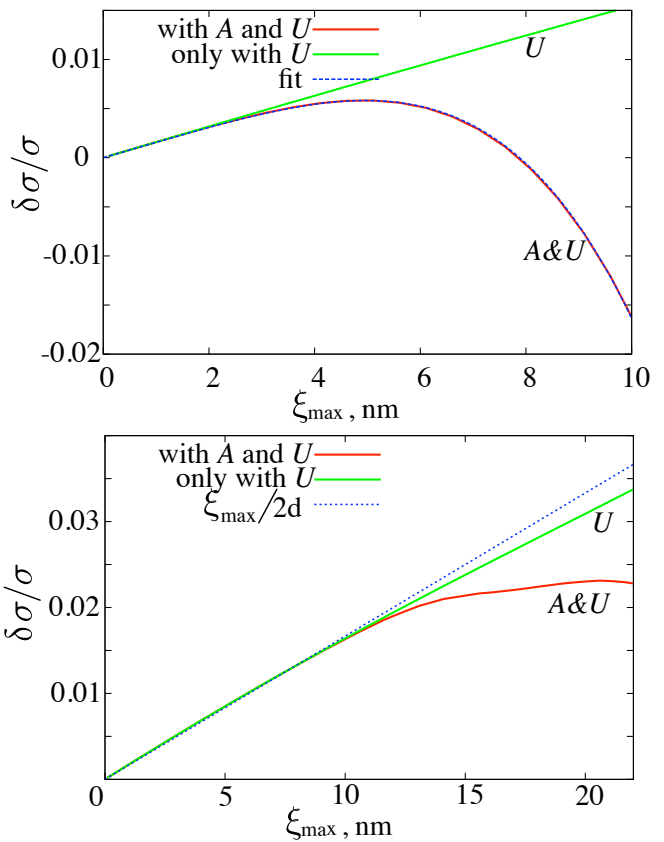}}
\caption{Dependence of the piezocorrection to conductivity on the maximum deviation, $\delta\sigma(\xi_{max})/\sigma$, at the fixed gate voltage $V_g=0.03$~V (top), $3$~V (bottom). We show both the correction related to the term due to non-uniform pseudovector potential, and the contribution without this term. For $V_g=0.03$~V, we also include the best fit $V = \alpha\xi_{max}+\beta\xi_{max}^4$ which represent the sum of linear in $\xi_{max}$ correction due to charge redistribution and the correction due to nonuniform pseudovector potential $\beta \xi_{max}^4$. Parameters of the graphene flake are~\cite{foot1}.}
\label{correctionAA}
\end{figure}

Fig.~\ref{correctiongate} shows the exact result of summation over modes Eq.~(\ref{correctionUmain}). At both high ($k_FL \gg 1$) and
low ($k_FL \ll 1$) gate voltages,
the correction follows the asymptotic behavior both for weak and strong residual stress $T_0$, Eqs.~(\ref{condt0}) and~(\ref{condth}).
On the same plot we compare the correction we found with the correction due to pseudomagnetic fields at the edges~\cite{Katsnelson}.
The latter one has a different sign (conductivity decreases with the stress).
For high residual stress this correction is more important than due to charge redistribution, according to Ref.~\onlinecite{Katsnelson} it can block conductivity.
For low residual stress it is about 10 times lower than the increasing conductivity correction.
The oscillations of $\delta \sigma/\sigma$ have the period of $k_FL$ d are associated with the shift of Fabry-Perot resonances in conductivity for deformed graphene flake as conpared with an undeformed flake. This shift occurs since the effective longitudinal wave vector of an electron in graphene depends on the deformation since it feels different charge density over the graphene flake. Note also that the contribution from pseudomagnetic fields does not oscillate since the value of $k_F$ is the same for the whole flake.
The first order perturbation theory in $\xi_{max}/d$ is valid until this parameter reaches a rather large value, $\xi_{max}/d\sim0.1$ (see Appendix for more details).

For the case of local deformation, using the graphene profile (\ref{localprofile}) and using the same technique as in Appendix, we find
the correction for conductivity due to the charge redistribution,
\begin{equation}
\frac{\delta\sigma}{\sigma} = \begin{cases}
\xi_{max}/2d, & \text{$1\ll k_FL;$}
\\
0.088\xi_{max}(k_FL)^2/d, & \text{$k_FL\ll
1.$} \end{cases}
\label{correctionAFM}
\end{equation}

Note that the asymptotic behavior for large $k_FL$ has the same form as for homogeneous deformation.

We can also estimate the influence of inhomogeneous pseudomagnetic field assuming the local form of Hooke's law as in~\cite{Isacsson}
and using the perturbation theory for the transfer matrix, as detailed in Appendix. We find that the first order perturbation theory correction in pseudovector potential vanishes, whereas the second order correction can be estimated as
\begin{equation} \frac{\delta \sigma_A}{\sigma}\approx5.5 \cdot 10^5\frac{\xi_{max}^4}{L^4} \sqrt{\frac{d[\mu\text{m}]}{V_g [\text{V}]}}.\label{secondcorrection}\end{equation}
At low gate voltages and large deformations (for instance, induced by local deformation), this correction can be more important that the one from the charge redistribution, and thus the conductivity will be suppressed. From comparison of Eqs.~(\ref{secondcorrection})~and~(\ref{correctionAFM}) this supression happens for deformations:
\begin{equation} \frac{\xi_{max}}{L}>10^{-2}\sqrt[6]{\frac{d[\mu\text{m}]}{V_g [\text{V}]}}\left(\frac{L}{d} \right)^{1/2}\label{suppression}.\end{equation}
We demonstrate this by solving numerically by transfer matrix method the Dirac equation with additional potential due to charge redistribution and pseudovector potential, Fig.~\ref{correctionA2}. At fixed large $\xi_{max}=12$~nm (estimated using Eq.~(\ref{suppression})) and at low voltages the conductivity starts to decrease due to inhomogeneous tension distribution in the flake, and at higher voltages increases again due to the effect of charge redistribution. Fig.~\ref{correctionAA} shows that for small gate voltages lower maximum deformation $\xi_{max}$ is  required to reach the point where the conductivity starts to decrease, in agreement with Eq.~(\ref{secondcorrection}). For high gate voltages $V_g\sim 3$~V pseudomagnetic fields lead to saturation of the conductivity rather than to its decrease.

\subsection{Two-gate geometry}
\label{subsec:twogates}

\begin{figure}[tb]
\centerline{\includegraphics[width=\linewidth]{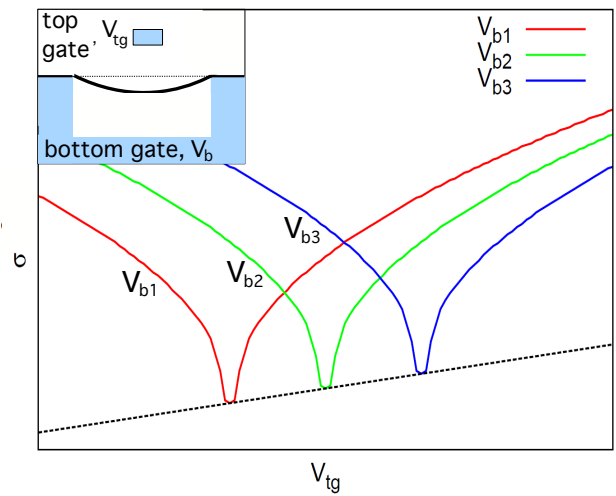}}
\caption{Schematic dependence of conductivity of suspended graphene on the top gate voltage for several fixed bottom gate voltages. The conductivity at the Dirac point is slightly shifted due to change in Fermi velocity caused by deformation. The difference between the values of conductivity of the Dirac peaks for different bottom gate voltages, $V_{b1}$ and $V_{b2}$, is proportional to the difference in relative deformations, $(\sigma_D(V_{b1})-\sigma_D(V_{b2}))/\sigma_D\sim u_{xx}(V_{b1})-u_{xx}(V_{b2})$.}
\label{topgate}
\end{figure}

Conductivity can also be used to measure relative stretching of deformed suspended graphene. Note that the influence of stretching on the conductivity of graphene deposited on a substrate has been demonstrated experimentally~\cite{grow3}. For suspended graphene it is more difficult to extract the value of stretching than from the graphene on the substrate, since the gate voltage simultaneously varies the concentration and deforms graphene, as shown above.

To measure relative stretching of suspended graphene, we propose the two-gate geometry (Fig.~\ref{topgate}). The deformation of the graphene flake is created by the large bottom gate, the influence of the top gate on the deformation is small as the top gate is narrow. The top gate is used to vary the charge density in the region underneath it. In this geometry at the fixed voltage at the bottom gate one can move through the Dirac point by varying the voltage at the top gate (the experiment for bilayer with two gates on the substrate~\cite{Oostinga}). The value of conductivity at this point depends on the deformation.

The stretching of the graphene flake, as discussed above, induces variations of the conductivity for two reasons. First, it induces pseudomagnetic fields. These, however, can be gauged away of Dirac equation \cite{Ludwig} at the Dirac point and do not influence the conductivity. Second, it shifts the Fermi velocity. The relative shift is proportional to the deformation, $\delta v_F/v_F\sim\xi_{max}^2/L^2$, and leads to the positive correction of the conductivity at the Dirac point, $\delta\sigma/\sigma\sim\delta v_F/v_F\sim\xi_{max}^2/L^2$. Thus, for different bottom gate voltages, which is equivalent to different maximum deformations $\xi_{max}$, the conductivity at the Dirac point is slightly different, and the relative graphene stretching can be restored from this dependence. For example, consider the dependence of conductivity on the top gate voltage for different bottom gate voltages (Fig.~\ref{topgate}).
At a fixed value of the bottom gate voltage, the conductivity as a function of the top gate voltage exhibits a peak dependence, with the minimum corresponding to the Dirac point. The difference between the values of conductivity at Dirac peaks, $\sigma_D$, for different bottom gate voltages, $V_{b1}$ and $V_{b2}$, is proportional to the difference in relative deformation, $(\sigma_D(V_{b1})-\sigma_D(V_{b2}))/\sigma_D\sim u_{xx}(V_{b1})-u_{xx}(V_{b2})$ (we remember that $u_{xx}\sim \xi_{max}^2/L^2$).

\section{Discussion} \label{sec:discussion}

In this Article, we investigated two mechanisms which affect the conductivity of suspended graphene --- charge redistribution induced by the gate(s), and  pseudomagnetic fields induced by the deformation of graphene. We find that for the small residual stress $T_0$, the charge redistribution mechanism dominates. For low gate voltages and strong deformation, which experimentally is best realized by using AFM, the correction due to nonuniform pseudomagnetic fields is more significant. The correction due to pseudovector potential at the region of suspension can decrease conductivity at the large residual stress~\cite{Katsnelson}. It is important that the two mechanisms provide corrections to conductivity which are of different signs. Indeed, the correction from pseudomagnetic fields suppresses the conductivity~\cite{Katsnelson} by shifting K-points due to the vector potential. The shift is different at different points of the suspended sample, and if the deformation is big enough, the Fermi circles at the clamping points and at the centre of the flake do not overlap: The system becomes insulating. If now we take into account the effects of the gate, not only the Fermi circles are shifted, but their radii are greater at the center of the flake since the charge density is greater in the areas closer to the gate. The increase of the radii and the shift of the center compete, and we find that typically the radius increase is more important.

It is difficult to measure piezoconductivity only by using a bottom gate since the gate voltage not only bends graphene and produces the correction to the conductivity, but also shifts the overall charge density. The density dependence of the conductivity is different from the density dependence of the correction. Thus, to extract the value of piezoconductivity, one has to compare the conductance of deformed and undeformed graphene sheet at the same density, which can only be done in the one-gate geometry by comparing the results with the theoretical prediction. In contrast, the two-gate setup, with a bottom gate fixing the overall density and the top gate (which can be an AFM tip) creating the deformation is more convenient to extract piezoconductivity. One can fix the voltage on the bottom gate and start to deform the flake with the AFM tip. At low gate voltages the conductivity decreases due to the pseudomagnetic fields, whereas at higher voltages it starts to grow due to the charge redistribution.

In the real experimental situation, the AFM tip has a point shape, whereas in this Article we considered for illustration the deformation homogeneous in one direction, {\em i.e.} replaced the tip by a rod. Non-homogeneous deformation in all directions creates pseudomagnetic fields, with the conductivity depending not only on the transverse displacement, but also locally on the position over the graphene sheet. The conductivity is the largest if the tip is placed in the middle of the sheet, and decreases if the tip moves to the side. We can understand this behavior from a simple reasoning. Indeed, the electrons which from the two sides of the tip feel the pseudomagnetic fields and interfere similarly to an Aharonov-Bohm ring. The interference is more destructive if the tip is further from the center, and thus the conductivity decreases.

Another parameter which affects the conductivity is the residual stress $T_0$. It can be varied experimentally for instance if one uses graphene suspended over piezosubstrate. Putting voltage on the substrate would induce extra stress on graphene, and one can move from the situation where pseudovector potential blocks the conductivity at low gate voltages to the case where residual stress does not play a role and the correction due to charge redistribution increases the conducitivity.

In this Article, we considered ideal ballistic graphene. In particular, we disregarded the contact resistance, assuming the clamping points to be ideal contacts. Finite transparency of the contacts would suppress both the conductivity itself and the piezocorrection to the conductivity; in addition, it would raise the amplitude of Fabry-Perot resonances.

For strong deformations of the graphene sheet, the problem becomes much more complicated, since one has now to solve elasticity equations self-consistently, taking into account that the displacement depends on the charge redistribution. This leads to additional terms in the equations of the elasticity theory. Taking into account influence of the density redistribution on the term with electrostatic pressure in the equation of deformation, one can show that the self-consistency condition increases the deformation in the middle of the graphene sheet. This effect only becomes important at sufficiently strong deformations.

Finally, we assumed that undeformed graphene is flat. In reality, it is always rippled, and, in principle, one needs to use the elasticity theory for membranes. However, we do not expect that taking ripples into account would significantly affect the results of this paper. First, the ripples are small and have a large radius of curvature, which means they are very little affected by the overall deformation of the graphene sheet. Second, the main effect of the ripples is to renormalize the energy over the graphene sheet~\cite{Ostrovsky}. We thus expect that our results are valid, but for renormalized energy over the flake (energy is determined by gate voltage in clean case, and is renormalized in the rippled case).

\section{Acknowledgements}

We thank A. Akhmerov, C. Beenakker, M. Fogler, S. Goossens, F. Guinea, A. Isacsson, and A. Kharimkhodjaev for discussions.
We acknowledge the financial support of the Future and Emerging Technologies program of the European Commission, under the FET-Open project QNEMS (233992), of the Dutch Science Foundation NWO/FOM and of the Eurocores program EuroGraphene.

\section{Appendix. Perturbative corrections to conductivity}

In this Appendix, we calculate the corrections to the conductivity due to both charge redistribution and pseudomagnetic fields, using the preturbation theory.

The Dirac equation for one valley in graphene has the form
\begin{equation} v_F \vec{\sigma}\vec{p} +\delta U(x,y)=E \ , \label{diracgeneral}\end{equation}
with $\vec{\sigma}=(\sigma_x,\sigma_y)$, $\vec{p}=(p_x,p_y)$,
$$p_x=-i\hbar\partial_x+A_x, p_y=-i\hbar\partial_y+A_y,$$
$A_x(x,y)$ and $A_y(x,y)$ being the components of the pseudomagnetic vector-potential~\cite{Cbeta}, given by Eq. (\ref{pseudovector}), and $\delta U(x,y)$ is the  additional electrostatic potential due to the charge redistribution over the graphene flake. It is determined by local variations of the Fermi energy over the flake. Since the Fermi energy depends on the charge density over the flake, $E_F(x)=\hbar v_F k_F(x)$, $k_F(x)=\sqrt{\pi n(x)/e}$, one has
$$\delta U(x)/E=\delta k_F(x)/k_F=\delta n(x)/2n=\xi(x)/2d.$$
\begin{figure}[tb]
\centerline{\includegraphics[width=\linewidth]{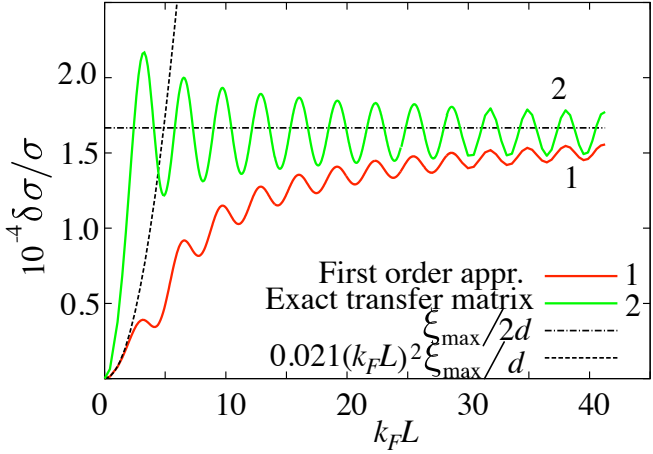}}
\caption{Dependence of the relative correction to conductivity on $k_FL$,
$\delta\sigma/\sigma(k_FL)$ for constant $\xi_{max}/d=1/3000$, for the correction of the first order in
$\xi_{max}/d$, Eq.~(\ref{correctionU}), the curve $1$, and exact transfer matrix solution
of integral equation, the curve $2$.
The correction from the exact solution has the same dependence on $k_FL$
as the first order correction, the oscillations are in the same phase.
Asymptotics for small and large $k_FL$, Eq.~(\ref{correctionUmain}), are
shown as dashed lines. Parameters of the flake are~\cite{foot1}.}
\label{exactkfl}
\end{figure}
We only consider the deformation homogeneous in $y$-direction.Then both $A_y$ and $\delta U$ only depend on the coordinate $x$, and $A_x=0$ (see Section~\ref{sec:piezo}). The problem becomes effectively one-dimensional since the momentum $q$ in $y$-direction is conserved. It is convenient to use the transfer matrix representation of Dirac equation~\cite{Ostrovsky} to calculate the correction to the conductivity caused by the deformation $A_y(x)$, $\delta U(x)$,
\begin{eqnarray} &\mathcal{T}_H&(x_2,x_1) = \mathcal{T}_{0H}(x_2,x_1)- \\
& -& \int_{x_1}^{x_2} dx
\mathcal{T}_{0H}(x_2,x)\left(\sigma_z \delta U(x)+i\sigma_x A_y(x)\right) \mathcal{T_H}(x,x_1)\ ,\nonumber \label{transfer}
\end{eqnarray}
where $\mathcal{T}_H$ is the Hadamard transformed transfer matrix, and $\mathcal{T}_{0H}$ is the Hadamard transformed transfer matrix of the unperturbed system,
\begin{equation} \mathcal{T}_{0H}=\exp\left(i\sigma_z k_F L+\sigma_x qL\right).\end{equation}
We perform the perturbation expansion of the integral form for Eq.~(\ref{transfer}), and in the first order in $\delta U(x)$ and $A_y(x)$ we obtain
\begin{eqnarray} \mathcal{T}_1(x_2,x_1)=\mathcal{T}_0(x_2,x_1)-
\\-i\int_{x_1}^{x_2} dx
\mathcal{T}_0(x_2,x)\left(\sigma_z \delta U(x)+i\sigma_x A_y(x)\right) \mathcal{T}_0(x,x_1). \label{transferfirst}\nonumber
\end{eqnarray}

The conductance of the graphene sheet is determined by Landauer formula~(\ref{Landauer}).
According to general scattering theory~\cite{Ostrovsky}, the transmission matrix element $\hat t$ is an inverse element of $\mathcal{T}_{H}$,
\begin{equation} \hat t=\left(\mathcal{T}_H^{--}\right)^{-1}.\label{t}\end{equation}
Taking into account Eq.~(\ref{transferfirst}), Landauer formula~(\ref{Landauer}), and the definition~(\ref{t}), the first order corrections to conductivity due to electrostatics and pseudo-magnetic field are
\begin{eqnarray} \delta\sigma_{U}&=&\sum_q 4 \left\vert t_q \right\vert^4 I_U k_F L \frac{q^2 k_F}{k^3} \sin kL, \label{correctionU}\\
I_U&=&\int_{0}^{L} \frac{d x}{L} \frac{\xi(x)}{2d} \sin k(L-x) \sin kx,\nonumber\\
\delta\sigma_{A}&=&
\sum_q 2\left\vert t_q\right\vert^{4}
\frac{k_F q }{k^2}I_A\label{correctionA},\end{eqnarray}
\begin{eqnarray} I_A&=&\int_{0}^{L} d x \delta
A(x)\times\nonumber\\
&\times&(\sin^2 k L - 2\cos k L \sin k x \sin k (L-x)),\end{eqnarray}
where $q=2\pi n/W$ is a wave vector in the $y$-direction, $n$ is an integer number, and
$k$ is a wave vector along the strip, $$k^2+q^2=k_F^2.$$
Furthermore, $t_q$ is the transmission probability for clean system for the mode $q$, and $$\left \vert t_q \right \vert ^2=(\cos^2 kL + k_F^2 \sin^2 k L/k^2 )^{-1}.$$
Note that the first-order correction due to the pseudo-vector potential (\ref{correctionA}) only contains odd powers of $q$, so that the sum over $q$ vanishes. Thus, the first-order correction to the conductivity is determined solely by the density redistribution. It is linear in the maximum deviation $\xi_{max}/d$ for small deviations.

\begin{figure}[tb]
\centerline{\includegraphics[width=\linewidth]{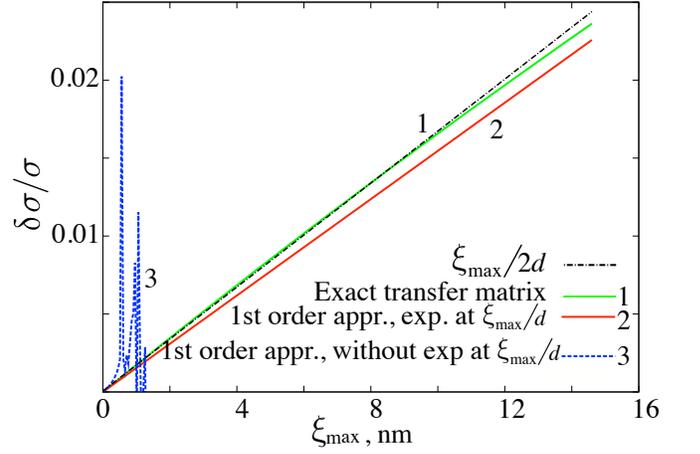}}
\caption{Dependence of the relative correction to conductivity on
the maximum deformation, $\delta\sigma(\xi_{max})/\sigma$ for constant $k_FL=40$, in the first order in
$\xi_{max}/d$, Eq.~(\ref{correctionU}), the curve $2$, and the exact transfer matrix solution
of integral equation, the curve $2$.
Both the expansion and summation of the $(\mathcal{T}^{--})^{-1}$ are not
valid for $\xi_{max}k_FL/d>1$ for small parameter as mentioned in the
text. The dashed line, the curve $3$, is the summation result. The exact solution shows
linear dependence on deviation even for rather large deviation, and this
linear dependence is close both to the correction Eq.~(\ref{correctionU})
and to the correction averaged over fast oscillations,  $\xi_{max}/2d$. Parameters of the flake are~\cite{foot1}.}
\label{exactxi}
\end{figure}

First, we remark on the validity of Eq.~(\ref{correctionU}). The expansion of the expression
$$1-4(t_qt_q^{\dagger})^2 I_U k_F L \frac{q^2 k_F}{k^3} \sin kL$$
has been made under assumption that
the second term is small in comparison with unity due to the small
prefactor $\xi_{max} k_F L/d$. Following this argument, the expression for the first order correction to the conductivity in
$\xi_{max}/d$ is formally only valid for $\xi_{max}/d
k_F L\ll 1$. However, solving the integral equation numerically, we find that this expression
is valid for a broader parameter range. We compare results of calculations for the first
order correction Eq.~(\ref{correctionU}) and numerical solution of
Eq.~(\ref{transfer}) for the two cases: for the fixed ratio $\xi_{max}/d$ and for
the fixed value of $k_FL$. For the first case, the dependence of $\delta
\sigma/\sigma$ on $k_FL$ shows the same oscillation period and the same
asymptotic behavior at large $k_FL$, Fig.~\ref{exactkfl}. For the second
case, at large $k_FL \sim 40$ (for the distance to the gate $d=300$ nm this corresponds to the gate voltage
$V_g=3$ V) the expansion clearly ceases to be valid, see Fig.~\ref{exactxi}.
We thus conclude from the results of our numerical solution that
the expression for the correction linear in $\xi_{max}/d$ is applicable until
$\xi_{max}/d\ll 1$, which is weaker than the perturbation theory suggestion $\xi_{max} k_F L/d\ll 1$.

The second order correction to conductivity contains also a term with the
pseudo-vector potential, the magnitude of the term being
$(\xi_{max}/L)^{4}$.

We consider both corrections separately.
Now we perform the analysis of Eq.~(\ref{correctionU}) for
deformation with constant pressure. For this case, the shape of the
strip is nearly parabolic (Section~\ref{sec:deformation}) and can be
approximated as
$$ \xi(x)= \frac{4\xi_{max}}{L^2} (x-L/2)^2. $$
The integral with the induced potential $\delta U(x)$ from
Eq.~(\ref{correctionU}), $I_U$, is
\begin{equation} I_U=
\frac{\xi_{max}}{12 d}\frac{kL(6-(kL)^2)\cos kL -3 (2-(kL)^2)\sin
  kL}{(kL)^3}\label{sinint}.\end{equation}
Now we can perform the summation over modes for $\delta \sigma$,
Eq.~(\ref{correctionU}), analytically in two asymptotic cases: $k_FL\ll1$
and $k_FL\gg1$.

For $k_FL\ll1$, the evanescent modes give the most important contribution
to the conductivity~\cite{BeenakkerR},
\begin{equation} \sigma(k_FL\ll 1)=\frac{W}{2\pi
L}\int_{-\infty}^{\infty}\frac{dx}{\cosh^2 x} = \frac{W}{\pi L},\end{equation}
and to the correction to the conductivity,
\begin{equation} \delta \sigma (k_FL\ll1)= \frac{\xi_{max}}{3d}\frac{W}{2\pi L} (k_F
L)^2 I\ ,\end{equation}
with $$I=2\int_{0}^{\infty}\frac{dx \sinh x (x(6+x^2)\cosh x-3(2+x^2)\sinh
x)}{x^4\cosh^4 x},$$
and its numerical value is $I\approx 0.124$.
The relative correction to conductivity reads
\begin{equation} \frac{\delta \sigma}{\sigma} =
\frac{\xi_{max}I}{6d}(k_FL)^2\approx0.021 \frac{\xi_{max}}{d}(k_FL)^2.\end{equation}

For $k_FL \gg 1$ we average over fast oscillations.
In this case, only the propagating modes contribute significantly to
the conductivity.

To perform the averaging, we replace the summation over $q$ by the integration,
$$\sum_q\longrightarrow \frac{W}{2\pi} \int dq.$$
To simplify subsequent calculations, we make the change of variables
$q=k_F\sin \phi$,
$k=k_F\cos\phi$, and then go from the integral over $dq$ to the integral over $d\phi$.
The correction to the conductivity Eq.~(\ref{correctionU}) has the form
\begin{eqnarray} &\delta \sigma &= \frac{k_FW}{\pi}\frac{\xi_{max}}{d}
\times \nonumber \\
  &\times& \int_0^{\pi/2}d\phi \frac{\cos\phi \sin^2\phi
  \sin^2(k_F L\cos\phi)}{(\cos^2\phi\cos^2(k_F L\cos\phi)+\sin^2(k_F
L\cos\phi))^2}. \nonumber
\end{eqnarray}
In this expression only the term with $3 \sin kL/ kL$
from~Eq.(\ref{sinint}) survived:
All terms with $\cos kL$ vanish after averaging, and the term with $-6 \sin
kL/ (kL)^3$ is smaller
than one which is taken into account).
For large $k_FL$ the terms $\cos (k_F L \cos\phi) $ and $\sin (k_F L \sin\phi)$
in Eq.~(\ref{correctionU}) oscillate very rapidly. We can represent
$\int_0^{\pi/2} d\phi$ as a sum  of fast oscillating terms, with each term
being an average over the period,
\begin{eqnarray}&\int_0^{\pi/2} d\phi \longrightarrow \nonumber \\
&\sum _{n=0}^{N_{max}} \int_{\phi_n}^{\phi_{n+1}} d\phi
f(\phi_{n+1/2},\cos(k_FL\cos\phi),\sin(k_FL\cos\phi)),\nonumber\end{eqnarray}
with $k_FL\sin\phi_n=2\pi n$. The integrand $f$ is determined by
structure of Eq.~(\ref{correctionU}) and Eq.~(\ref{sinint}),
\begin{equation} \int_0^{2\pi}f(\phi)=\int_0^{2\pi} \frac{\sin^2 x dx}{(a^2\cos^2
x+\sin^2 x)^2}=\frac{2\pi}{a}.\nonumber\end{equation}
What is left is the sum over $n$,
\begin{equation} \delta \sigma = \frac{\xi_{max}}{d} \frac{k_F W}{\pi} 2\pi
\sum_n\sqrt{1-(x_n/k_FL)^2},\end{equation}
$x_n=\pi(2n+1)$.
Finally,
\begin{equation} \delta \sigma=\frac{\xi_{max}}{8d} k_F W.\end{equation}
The conductivity after
averaging over fast oscillations becomes $\sigma=k_F W/4$,
and the relative correction to the conductivity is
\begin{equation} \frac{\delta\sigma}{\sigma}=\frac{\xi_{max}}{2d}. \nonumber \end{equation}
From general physical considerations about the correction (see main text),
one also expects the dependence $\delta \sigma/\sigma \sim \xi_{max}/d$ for $\delta \sigma/\sigma$.

Concerning the correction due to the pseudomagnetic fields, it is of the second order in $\delta A$, and the analytical expressions are too cumbersome.
Instead, we illustrate our conclusions using the numerical solution of the integral equation~(\ref{transfer}). It is done by multiplying transfer matrices for small intervals of the length $\delta x$. Convergence with the size of $\delta x$ is reached.


\begin{thebibliography}{99}
\bibitem{foot1} We consider the experimentally feasible parameters of the graphene flake: Length $L=1\mu m$, distance to the gate $d=300$ nm, Young modulus $E=1~$ TPa, Poisson ratio $\sigma=0.15$, and thickness of the graphene monolayer $h_0=0.34$nm.
\bibitem{Cbeta} Parameter $C\tilde{\beta}/a$ can be taken in the form $3 t_{str}\beta /k a$, where according to calculations~\cite{Woods} $k= 8.98 N/m^2$,
$\beta/a=0.4$ N/m$^2$, and the parameter $\beta/(ka)=0.0445$, $t_{str}$ can be determined by different methods, it is approximetely $t_{str}=2 \text{\AA}^{-1}$ (from femtosecond
time-resolved photoemission experiement~\cite{Hertel}, $t_{str}\approx
1 \text{\AA}^{-1}$, from analitycal estimations~\cite{Pietronero} and
optical spectrum of polyacetylene, $t_{str}\approx 2.0 \text{\AA}^{-1}$, from
tight-binding approximation $t_{str}\approx 2.5 \text{\AA}^{-1}$). Finally, the estimation of the
$C\tilde{\beta}/a$ is $0.27 \text{\AA}^{-1}$.
\bibitem{modelleads} Note that the conductivity in our considerations and in Ref. ~\onlinecite{Katsnelson} are calculated for different models of the leads. We use the model with infinitely doped leads like in Ref.~\onlinecite{BeenakkerR}, whereas in~\cite{Katsnelson} graphene has the same potential as the leads.
\end{thebibliography}
\end{document}